\documentclass[11p,a4paper]{article} 
\pdfoutput=1
\usepackage{jinstpub}

\usepackage{mathrsfs}
\usepackage{siunitx}
\usepackage{soul} 
\usepackage{lineno}
\modulolinenumbers[5]

\def \defaultplotwidth {0.6\linewidth}
\def \figwidth {.9 \linewidth}
\def \wideplotwidth {1.0\linewidth}

\title{Precision measurements of the scintillation pulse shape for low-energy recoils in liquid xenon}
\author[1]{E. Hogenbirk,\note{Corresponding author.}}
\author{J. Aalbers,} 
\author{P. A. Breur,}
\author{M. P. Decowski,} 
\author{K. van Teutem,} 
\author{A. P. Colijn} 

\affiliation{Nikhef and the University of Amsterdam, Science Park, 1098XG Amsterdam, Netherlands}
\emailAdd{ehogenbi@nikhef.nl}

\arxivnumber{1803.07935}

\abstract{
We present measurements of the scintillation pulse shape in liquid xenon for nuclear recoils (NR) and electronic recoils (ER) at electric fields of \num{0} to \SI{0.5}{kV/cm} for energies $<$\SI{15}{keV} and $<$\SI{70}{keV} electron-equivalent, respectively.
The average pulse shapes are well-described by an effective model with two exponential decay components, where both decay times are fit parameters.
We find significant broadening of the pulse for ER due to delayed luminescence from the recombination process.
In addition to the effective model, we fit a model describing the recombination luminescence for ER at zero field and obtain good agreement.
We estimate the best performance of a combined S2/S1 and pulse shape ER/NR discrimination and show that even with \SI{2}{ns} time resolution, the improvement over S2/S1 discrimination alone is marginal, so that pulse shape discrimination will likely not be useful for future dual-phase liquid xenon experiments looking for elastic dark matter recoil interactions.}


\notoc 

\begin{document}
\maketitle
\section{Introduction}
Over the past decade, the field of direct dark matter detection has been led by dual-phase xenon time projection chambers (TPCs) \cite{xe1t, lux, pandax}.
These experiments are particularly well-suited to investigate weakly interacting massive particles (WIMPs), the leading dark matter candidate.
The main challenge for these experiments is to reduce radioactive backgrounds to the level of a few events per year, so that an (almost) background-free environment can be achieved.
A crucial part of background rejection is the ability to distinguish electronic recoil (ER) events, caused by background gamma or beta radiation, from nuclear recoil (NR) events, which are expected from WIMP interactions.
In dual-phase xenon TPCs, this is usually achieved by using the ratio of the secondary scintillation (S2) to the direct scintillation signal (S1), which is a powerful discriminant.
Alternatively, there is considerable interest in using the scintillation pulse shape as a particle identification method. 
This method, called pulse shape discrimination (PSD), has been applied in, e.g., liquid organic scintillators \cite{organic} and in liquid argon, where it provides the main discrimination against the sizable $^{39}$Ar ER background \cite{ardm, deap}.
In xenon, the scintillation pulse shape has been shown to depend on ionization density (and therefore particle type and recoil energy) \cite{hitachi} and electric field \cite{kubota, dawson2005}, so that pulse shape discrimination is in principle possible.
However, due to the short timescales ($\sim$\SI{4}{ns} and $\sim$\SI{22}{ns} for the two exponential decay times for alpha particles and fission fragments \cite{hitachi}) compared to typical photomultiplier tube~(PMT) responses, pulse shape discrimination is challenging in liquid xenon detectors.
The method has nevertheless been successfully applied as a particle identification method, either in combination with the S2/S1 ratio \cite{kwong2010, luxpsd}, or by itself, i.e., in single-phase detectors \cite{dawson2005, ueshima2011, zeplin1, davies1994}.
In general, a single pulse shape parameter or an effective distribution (such as a single exponential) is used to describe the pulse shape.
Recently, the LUX collaboration measured the pulse shape by using a sum of two exponential functions with a fixed time constant~\cite{luxpsd}.
This distribution correctly describes the decay of the two excited states, but neglects any time delay in their production, caused primarily by recombination of electron-ion pairs.
This delay is non-negligible for $\mathscr{O}($\si{MeV}$)$ ERs, but the influence of recombination on the pulse shape was assumed to be minor based on extrapolations from measurements at high energy, as measurements at low energy are lacking~\cite{mock2014}.

This paper aims to directly determine the scintillation pulse shape for ER $<$\SI{70}{keV}$_{ee}$ and NR $<$\SI{15}{keV}$_{ee}$ at electric fields up to \SI{0.5}{kV/cm} using measurements from XAMS, a dual-phase TPC setup described in \cite{xams_paper}.
All measured energies are reported relative to gamma-ray (ER) calibrations, which we make explicit by using the unit \si{keV}$_{ee}$ (electron-equivalent) for both ER and NR. 
We use a sum of two exponential distributions to describe the pulse shapes, where all parameters are allowed to vary to capture the effect of recombination.
Additionally, we use a recombination model to describe the pulse shape for ER at zero field.
Using the pulse shape model extracted from the data, we calculate the theoretical improvement in discrimination when combining PSD and S2/S1 discrimination in future large-scale TPCs searching for elastic WIMP recoils.

\section{The process of scintillation} \label{sec:model}
\begin{figure}[h]
\begin{center}
\includegraphics[width=\figwidth]{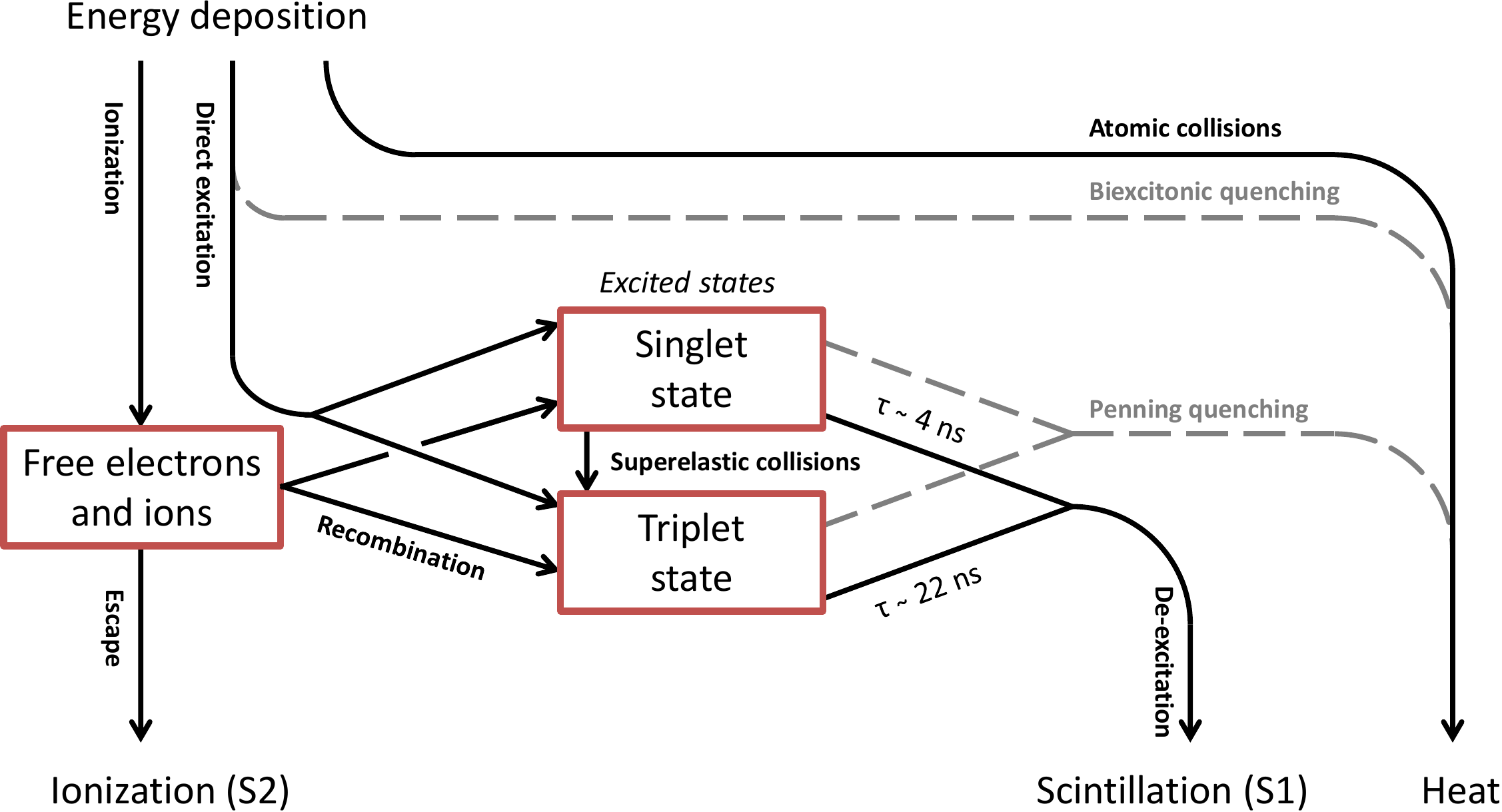}
\caption{After an energy deposition in liquid xenon, different processes lead to ionization, scintillation and heat. 
The processes indicated by the gray dashed lines are most relevant for recoils resulting in high ionization density.}
\label{fig:fig1}
\end{center}
\end{figure}

When ionizing radiation energy is deposited in liquid xenon, the energy is transferred to atomic excitation, ionization and heat (see figure~\ref{fig:fig1}). 
Excited xenon atoms combine with stable xenon atoms to form excited xenon dimers (Xe$_2^{*}$), which can be in either a spin singlet or triplet state.
The lifetimes of these states, $\tau_s$ and $\tau_t$, are measured at \num{2}~to~\SI{4}{ns} and $\sim$\SI{22}{ns}, respectively \cite{hitachi}.
The intensity of the scintillation light from the decay of one of these states is characterized by
\begin{equation} \label{eqn:2_1}
I_d(t, \tau) = \frac{1}{\tau} \exp{\left(\frac{-t}{\tau}\right)},
\end{equation}
with $\tau$ the lifetime of the state.
If we define the fraction of photons from the singlet state as $f_s$, we can write the time dependence of the direct scintillation as 
\begin{equation} \label{eqn:2_2}
I(t, \tau_{s}, \tau_{t}, f_s) = f_s I_d(t, \tau_{s}) + (1-f_s) I_d(t, \tau_{t}).
\end{equation}

In addition to the direct excitation, excited dimer states are formed by electron-ion recombination.
In this process, ionized electrons recombine with xenon ions to form doubly excited xenon atoms Xe$^{**}$, which finally form xenon dimer states.

The scintillation process in liquid xenon depends on the ionization density, which to first order is proportional to the linear energy transfer~(LET).
This has been studied extensively by using different LET tracks caused by (in order of increasing LET) high-energy ($\mathscr{O}($\si{MeV}$)$) ERs~\cite{keto, kubota1978, kubota}, low energy ($<$\SI{100}{keV}) ERs~\cite{xmass2016, luxpsd, akimov2002, ueshima_phd}, nuclear recoils~\cite{luxpsd, akimov2002, ueshima_phd}, alpha particles~\cite{hitachi, kubota1982} and fission fragments~\cite{hitachi}. 
First of all, the timescale of recombination increases for low LET tracks.
For high-energy ERs at zero field, the timescale is comparable to the decay times, so that the scintillation profile consists of the cumulative effect of dimer states forming and decaying.
In this case, a decay curve different from the distribution given in eq.~\eqref{eqn:2_2} has been found, with a single apparent decay constant of \SI[separate-uncertainty = true]{33 \pm 1}{ns}~\cite{keto, kubota1978} or $\sim$\SI{45}{ns} \cite{kubota} and significant broadening of the peak intensity.
Secondly, the scintillation yield decreases with LET, as there are quenching mechanisms that become more relevant for high ionization density tracks (indicated by the gray dashed lines in figure~\ref{fig:fig1}).
In biexcitonic quenching, two excitons are quenched before the excited molecular states are formed.
In Penning quenching, two excited molecular states collide to produce one excited state and one of the states dissociates to the ground state.
For nuclear recoils and fission fragments, the yield decreases further due to elastic energy transfer to xenon atoms, resulting in energy dissipation through atomic motion (heat) instead of scintillation.
Finally, the fraction of photons coming from the singlet state~$f_s$ increases with LET.
Measured values are \num{0.05} for $\mathscr{O}($\si{MeV}$)$ electrons~\cite{kubota1978}, \num{0.04} \cite{luxpsd} and \num{0.045} to \num{0.145} \cite{xmass2016} for low-energy ($<$\SI{100}{keV}) ERs, \num{0.21} for NRs~\cite{luxpsd}, \num[separate-uncertainty = true]{0.31 \pm 0.05}~\cite{hitachi} and $\sim$\num{0.6}~\cite{kubota1982} for alpha particles and \num[separate-uncertainty = true]{0.62 \pm 0.08} for fission fragments \cite{hitachi}.
It has been suggested~\cite{hitachi} that this is due to superelastic collisions, where a singlet excited state collides with a free electron and forms a triplet state.
As this requires a free electron, this process is more likely to occur if recombination is slow, which is the case for low energy transfer tracks.

The applied electric field influences the pulse shape due to its effect on free electrons.
If an electric field is applied, the ionization electrons are drifted away from the interaction site, separating the electrons from the ions that are pulled to the opposite side (but at a much lower drift speed \cite{xedet}), so that the recombination luminescence is suppressed. 
It should be noted, however, that even at high field of \SI{4}{kV/cm} an apparent lifetime of \SI[separate-uncertainty = true]{27 \pm 1}{ns} was found for electronic recoils \cite{kubota1978}, which disagrees with the $\sim$\SI{22}{ns} found for alpha and fission fragment recoils.

\subsection{Recombination model} \label{sec:recomb_model}
\begin{figure}[h]
\begin{center}
\includegraphics[width=0.9\linewidth]{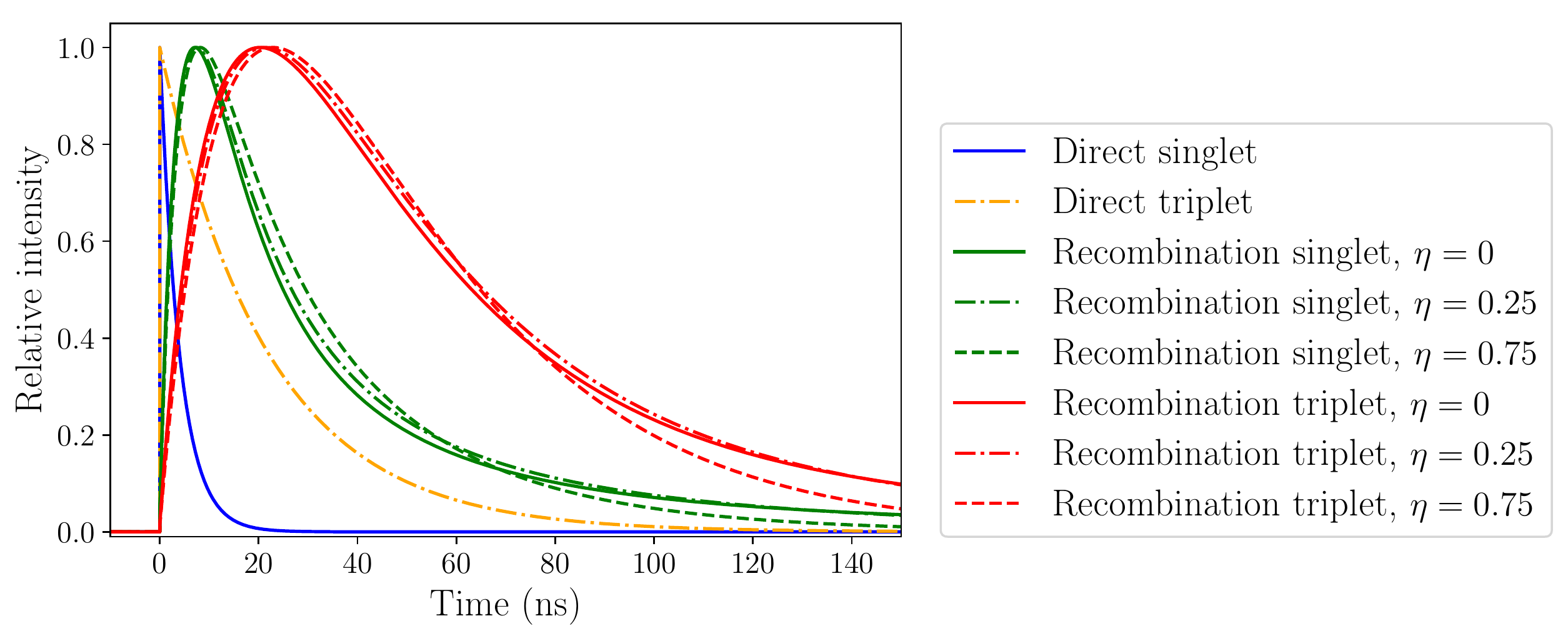}
\caption{Example curves of the scintillation time dependence components in eq.~\eqref{eqn:2_4}.
The time constants used are $\tau_s = $\SI{4}{ns}, $\tau_t$ = \SI{22}{ns} and $T_R$ = \SI{25}{ns}.
All curves are scaled such that the maximum is~1.
}
\label{fig:fig2}
\end{center}
\end{figure}

The time dependence of recombination luminescence in the absence of an electric field can be described following a model by Kubota et al.~\cite{kubota}.
This model assumes a region of uniform ionization density, where a fraction of the electrons~$\eta$ escapes at $t=0$ and the timescale of recombination is governed by the recombination time~$T_R$.
The derivation of the scintillation intensity time dependence is given in Appendix~\ref{app:rec} and yields
\begin{equation}
I_r(t, \tau, T_R, \eta) = A \exp{\left(\frac{-t}{\tau}\right)} \ \times  \int_0^t n_e(t')n_{ions}(t') \exp{\left(\frac{t'}{\tau}\right)}\ dt',
\end{equation}
with $A$ a normalization factor, $\tau$ the decay time of the excimer state, $T_R$ the recombination time and $\eta$ the probability of electrons to escape the recombination region.
The dependence on $T_R$ and $\eta$ is implicit through the solution of differential equations describing the electron and ion densities $n_e$ and $n_{ions}$.
For a general scintillation pulse, the time dependence in its most general form will be a combination of four terms, corresponding to the two states in the direct decay from excitations and the same two states formed by recombination:
\begin{eqnarray} \label{eqn:2_4}
I(t, \tau_s, \tau_t, T_R, f_s, f_R, \eta, f_s^R) = (1 - f_R) \cdot \left( f_s I_d(t, \tau_s) + (1 - f_s) I_d(t, \tau_t)\right) + \nonumber  \\ 
f_R \cdot ( f_s^R I_r(t, \tau_s, T_R, \eta) + 
(1-f_s^R) I_r(t, \tau_t, T_R, \eta)) ,
\end{eqnarray}
where the integral of $I$, $I_d$ and $I_r$ is normalized to~1 and $f_R$, $f_s$ and $f_s^R$ are the fraction of photons from recombination, the fraction of direct scintillation photons coming from the singlet state and the fraction of recombination photons coming from the singlet state, respectively.
Figure~\ref{fig:fig2} illustrates the time dependence of the scintillation intensity for the components of eq.~\eqref{eqn:2_4} for a specific choice of parameters. 

\subsection{Effective model}\label{sec:eff_model}
To avoid the complexity of including a recombination model and the uncertainty of the physical mechanisms, we introduce the effect of recombination by altering the lifetimes of the singlet and triplet states.
In this case, we follow the model of eq.~\eqref{eqn:2_2}, with the difference that the lifetimes are interpreted as effective lifetimes, i.e.,
\begin{equation} \label{eqn:2_5}
I(t, \tau_s^{\rm eff}, \tau_t^{\rm eff}, f_s) = f_s I_d(t, \tau_s^{\rm eff}) + (1-f_s) I_d(t, \tau_t^{\rm eff}).
\end{equation}
While eq.~\eqref{eqn:2_4} provides a model that may be motivated better physically, it comes at the cost of four extra parameters.
Therefore, the effective lifetime model can be chosen over the full model if only a working description of the pulse shape is required, or if the contribution from recombination is negligible.
In this paper, we use the model in eq.~\eqref{eqn:2_5} as an effective model and apply the model from eq.~\eqref{eqn:2_4} to the data at zero field only.

\section{Measurements}
\subsection{Data acquisition and calibration} \label{sec:calibration}
The data was collected with the XAMS setup, a dual-phase liquid xenon TPC.
The active volume of the TPC has a cylindrical shape with a diameter of \SI{44}{mm} and a total effective length (as defined by the distance between the cathode and gate meshes) of \SI{100}{mm}, giving a total xenon content of \SI{154}{cm^3} (\SI{434}{g} at \SI{-90}{\degree C}).
The walls of the TPC are made of Teflon to maximize the light detection efficiency. 
The active volume is viewed from below and above by two 2-inch circular PMTs (Hamamatsu type R6041-406~\cite{hamamatsu}).
For all the measurements described here, a voltage of \SI{3}{kV} is applied between the anode mesh and the gate mesh (\SI{5}{mm} separation), extracting the electrons from the liquid into the gas and providing the proportional scintillation.
The waveform data from both PMTs was collected by a CAEN V1730D digitizer with a time resolution of \SI{2}{ns}.
The high sampling rate, in combination with the relatively small geometry and the low transit time spread of the PMTs ($\sim$\SI{0.75}{ns}~\cite{hamamatsu}) makes the setup well-suited for pulse shape measurements.
Further details of the setup can be found in \cite{xams_paper, erik_master}.

\begin{figure}[h]
\begin{center}
\includegraphics[width=\defaultplotwidth]{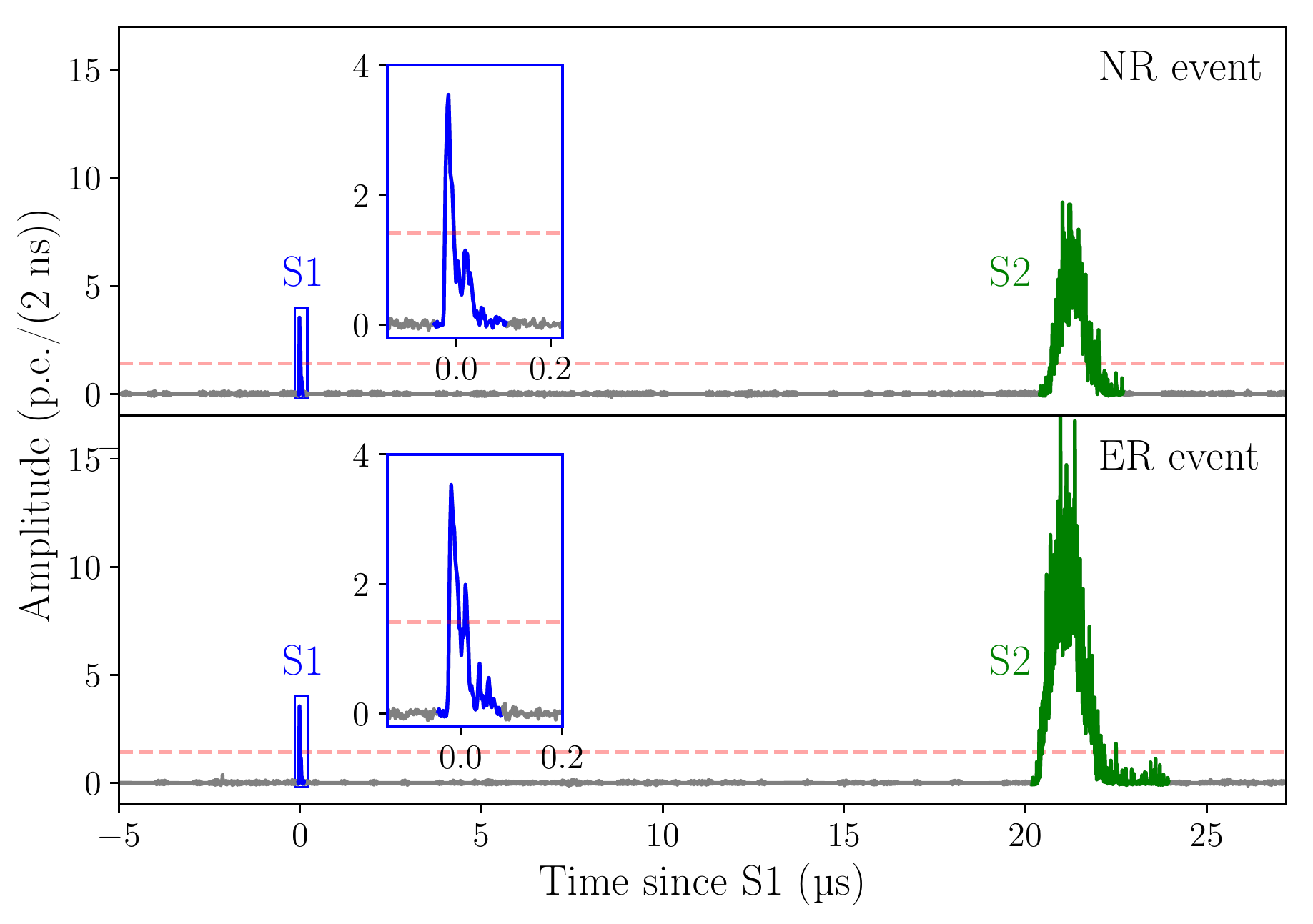}
\caption{Examples of the summed waveforms for an NR event (top panel) and an ER event (bottom panel) with comparable drift time. 
The inset shows a zoom of the S1 signal.
The dashed line shows the sum of the trigger levels for the two channels (note that a coincidence of both channels is also required).
The energy, reconstructed using a combination of S1 and S2, is \SI{9.5}{keV}$_{ee}$ and \SI{11.3}{keV}$_{ee}$ for the NR and ER event, respectively.
}
\label{fig:fig3}
\end{center}
\end{figure}

Figure~\ref{fig:fig3} shows the examples of the summed PMT waveforms for an NR and ER event.
The trigger was set at a level of \SI{7}{mV}, corresponding to \num[separate-uncertainty = true]{3.0 \pm 0.2} times the average single photoelectron~(SPE) amplitude for both channels, and the coincidence time window was set to the maximum of the on-board trigger of \SI{120}{ns}.
This allowed to trigger on small S2s, which is required for low-energy recoils.
Data processing followed the method outlined in \cite{xams_paper}.
Peaks were classified based on their width, which is defined as the time range containing the center 50\% of the area in the peak.
The peaks were classified as S1 if the width was less than \SI{60}{ns}.
For S2s, we required a width of at least \SI{100}{ns}, and an area of at least \num{100} times the average single photoelectron area.
In addition, for both S1 and S2 we required both PMTs to contribute to the signal.

An optical fiber connected to a blue pulsed LED was installed, allowing in-situ PMT calibration to SPE.
The gain of the PMTs was used to convert the area of the PMT signals to the number of photoelectron-equivalent, or p.e. (photo-electrons).
We use the photopeak in $^{137}$Cs and $^{22}$Na calibration data to correct the S1-signal for the z-dependent light detection efficiency (cS1) and the S2 for charge loss during electron drift (cS2).
For the S1, we fitted a second-order polynomial to the S1 as a function of~z.
All S1 values were then scaled according to this fit and normalized to the volume average.
The S2 correction was determined by an exponential fit to the photopeak as a function of drift time, where the normalization was to the point where the drift time is zero.
The electron lifetime was determined to be \SI[separate-uncertainty = true]{0.81 \pm 0.04}{ms}.
We calculate the recoil energy from
\begin{equation}
E = W\left( \frac{\rm cS1}{g_1} + \frac{{\rm cS2}_b}{g_2} \right),
\end{equation}
assuming a $W$-value of \SI[separate-uncertainty = true]{13.7 \pm 0.2}{eV} and using only the S2 in the bottom PMT (cS2$_b$). 
The photon gain $g_1 = (0.10 \pm 0.02)$~p.e./photon and electron gain $g_2 = (4.7 \pm 0.7)$~p.e./e$^-$ were obtained from the photopeak positions of $^{137}$Cs and $^{22}$Na, following the method described in \cite{dahl2009}.

For measurements with long timescales, we found that the S2 gain slowly decreased, while the S1-signal was unaffected. 
The gain decreased more rapidly during high-rate calibrations, and reset after cycling the anode power or temporarily raising the liquid level above the anode.
After ruling out faulty cabling or equipment, we attributed this to charge-up of floating Teflon dust particles, which we found in the detector after the measurements. 
These were probably produced while drilling a hole in the TPC wall to install the optical fiber.
We corrected for the S2 decrease by fitting the electronic recoil band in half-hour time slices; the correction is typically \SI{15}{\%} and at most \SI[separate-uncertainty = true]{30}{\%} (after \SI{3}{hours}). 

To collect the electronic recoil data, we acquired data with $^{137}$Cs (\SI{662}{keV} gamma), $^{22}$Na (positron decay, where we trigger on the \SI{511}{keV} back-to-back gammas using an external NaI(Tl) detector) and background radiation with \SI{0.5}{kV/cm} field, and with $^{137}$Cs only with \SI{0.1}{kV/cm} and zero field.
We combined all three datasets ($^{137}$Cs, $^{22}$Na and background) at \SI{0.5}{kV/cm} into one ER dataset to increase statistics.
For the nuclear recoil data, a $^{241}$AmBe neutron source with a neutron intensity of \SI{1.3e3}{n / s} was used.
A lead shield of \SI{25}{mm} was used to attenuate the gamma radiation from $^{241}$Am.
In addition, a cylindrical lead shield with a thickness of \SI{12}{mm} was placed outside the outer cryostat vessel to reduce the influence of background radiation for all measurements.

\subsection{Data selection} \label{sec:data}
For all data taken at nonzero drift fields we require exactly one S1 and one S2, reducing pileup and double scatter events, respectively.
To further reduce the number of double scatter events, we remove events with a wider S2 signal than expected based on a diffusion model fit to calibration data, yielding a diffusion constant of \SI[separate-uncertainty = true]{36.9 \pm 0.6}{cm^2/s} and \SI[separate-uncertainty = true]{15.5 \pm 0.6}{cm^2/s} for a drift field of \SI{0.1}{kV/cm} and \SI{0.5}{kV/cm}, respectively.
These cuts combined give an estimated double scatter rejection resolution of \SI{2}{mm} in depth.
Furthermore, we cut events occurring within \SI{1}{ms} of a previous event.
This removes events that triggered on a tail of single-electron S2s originating from photoionizations caused by large events.
Finally, we cut events within the top and bottom \SI{5}{mm}, giving an effective mass of \SI{390}{g} of liquid xenon in the TPC.

Only an S1-signal is generated for the data taken with zero field.
Nevertheless, the z-coordinate of the interaction can be estimated using the distribution of light across the two PMTs, which we related to the z-position using calibration data taken at nonzero drift field.
We cut on the z-coordinate to remove energy depositions above the gate, in the gas layer or below the cathode.
In addition, we used the z-coordinate to correct for the position-dependent light detection efficiency described in section~\ref{sec:calibration}.
This method breaks down for small S1s ($\lesssim$\num{40}~\!p.e.) because of statistical fluctuations in the light distribution.
We therefore set a lower energy limit of \SI{10}{keV}$_{ee}$.
The S1 classification was relaxed to include all peaks where both channels contribute.
Events with more than one S1-peak were still discarded, and the holdoff of \SI{1}{ms} from the previous event was also still applied to ensure a stable baseline condition.

\subsection{Electronic and nuclear recoil selection}
\begin{figure}
\begin{center}
\includegraphics[width=\wideplotwidth]{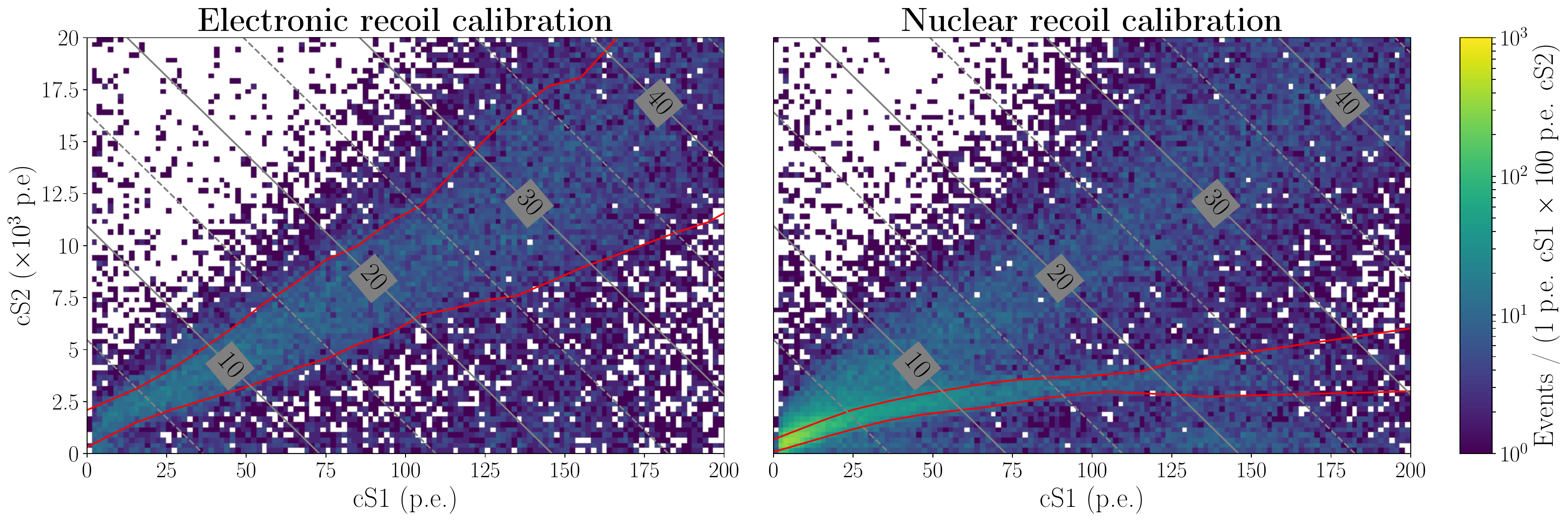}
\caption{Event distributions in (cS1-cS2)-space for electronic recoil (ER, left panel) and nuclear recoil (NR, right panel) calibrations.
Note that the density color scale is logarithmic.
The red lines indicate the fitted ER and NR 1$\sigma$ bands.
The gray lines indicate the reconstruced recoil energy, with the labels in \si{keV}$_{ee}$.
In the NR calibration, the ER band is of a similar magnitude as observed in background data taking.
}
\label{fig:fig4}
\end{center}
\end{figure}
The electronic and nuclear recoil events were selected based on the charge-to-light ratio cS2/cS1.
Figure~\ref{fig:fig4} shows the event distribution for the combination of all ER data (left panel) and the NR data (right panel) for a field strength of \SI{0.5}{kV/cm}.
In the nuclear recoil calibration, a distinct NR band appears below the ER band.
As the rate of nuclear recoils is relatively low, an ER band is still present in the nuclear recoil calibration dataset due to background radiation.
We observe some leakage of events below the bands, which we attribute to events at the edges of the detector, where there is incomplete charge collection and therefore a reduced S2 size.
This is consistent with the lower S2 width observed for these events, coming from the shorter gas gap due to the capillary effect of liquid xenon.

\begin{figure}
\begin{center}
\includegraphics[width=\wideplotwidth]{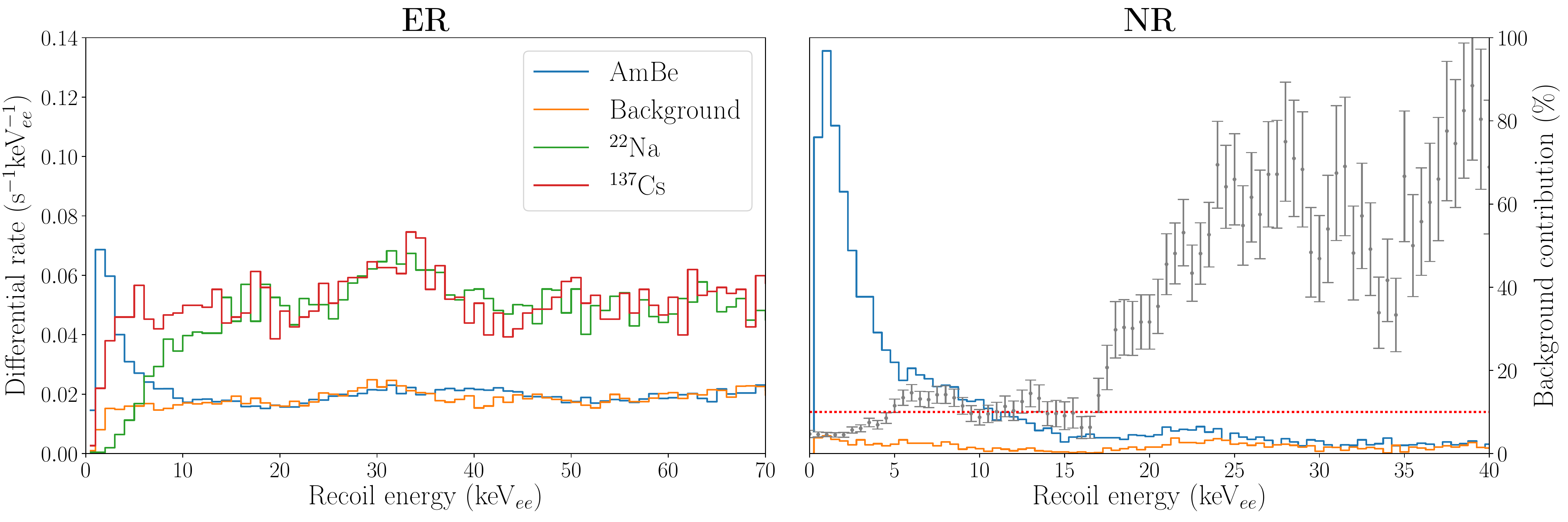}
\caption{
The energy spectrum for the ER (left panel) and NR (right panel) selection shown in figure~\ref{fig:fig4} for different sources.
\emph{Left panel: }
The AmBe and background ER rate are consistent above \SI{10}{keV}$_{ee}$, where the bands are separated, except for a small excess at $E=$~\SI[separate-uncertainty = true]{40 \pm 2}{keV}$_{ee}$ from the inelastic transition of $^{129}$Xe.
A broad peak at \SI[separate-uncertainty = true]{31 \pm 2}{keV}$_{ee}$, caused by xenon X-rays, is visible in all ER spectra.
The low-energy rolloff of the $^{22}$Na spectrum comes from the lower efficiency of the S1-only trigger.
\emph{Right panel: }
Energy spectrum for all events within the NR selection in AmBe and background data.
The gray points and error bars indicate the ratio between background and AmBe data, which gives the ER contribution in the NR dataset.
These points have been smoothed by a three-point moving average filter to reduce the statistical fluctuations at energies $\gtrsim$ \SI{25}{keV}$_{ee}$.
At energies below \SI{15}{keV}$_{ee}$, the background contribution is at roughly 10\%, as indicated by the dotted red line.
The broad peak observed in neutron and background data between \num{15} and \SI{35}{keV}$_{ee}$ is due to leakage of the \SI[separate-uncertainty = true]{31 \pm 2}{keV}$_{ee}$ xenon X-ray line below the ER band.
}
\label{fig:fig5}
\end{center}
\end{figure}

The ER and NR bands are fit by slicing the data in cS1 and performing an unbinned Gaussian fit to the distribution of cS2 in each slice.
The fit range is determined iteratively; an initial fit is done, after which the fit range is adjusted to $\mu \pm 1\sigma$ of this fit.
This procedure is repeated until convergence is reached.
The resulting ER and NR bands are indicated in the left and right panel of figure~\ref{fig:fig4}, respectively.
ER and NR data selection is consequently performed by cutting all the events outside the $\pm 1\sigma$ lines indicated in figure~\ref{fig:fig4}.

Figure~\ref{fig:fig5} (left) shows the ER spectra for three different sources and background data.
The rate of events in the ER band in the AmBe dataset is consistent with the background rate (which we can confirm above \SI{10}{keV}$_{ee}$ where the separation between the bands is good enough), except for a small excess of events at $E=$~\SI[separate-uncertainty = true]{40 \pm 2}{keV}$_{ee}$.
This is attributed to the neutron-induced inelastic transition of $^{129}$Xe, which gives a gamma ray at \SI{39.6}{keV}. 
The appearance of this excess also confirms our calibration of $g_1$ and $g_2$ at the low energies of interest.
Another broad structure was found in all the datasets at \SI[separate-uncertainty = true]{31 \pm 2}{keV}$_{ee}$.
We believe this is due to an X-ray energy transition from xenon at \SI{29.8}{keV} \cite{xray}.
If a photoelectric absorption interaction occurs in an insensitive region of the detector, an X-ray can result and penetrate into the sensitive volume.
This hypothesis is strengthened by the observation that the excess increases in intensity with higher ER rate.
Also, we find significantly more leakage to low S2 values at this energy, indicating that the events are concentrated at the edges of the detector, where there is incomplete charge collection.
This is consistent with the low range ($\sim$\SI{0.4}{mm}) of X-rays at this energy~\cite{nist}.
For the background and $^{137}$Cs data (which have the same trigger), we evaluate the low-energy acceptance by extending the observed flat spectrum between 5 and~\SI{20}{keV}$_{ee}$ to lower energies.
This gives an acceptance of~\num[separate-uncertainty = true]{0.7 \pm 0.1} at \SI{2}{keV}$_{ee}$ and reaching $>$\num{0.9} at \SI{3}{keV}$_{ee}$.
For the $^{22}$Na data taking, the trigger was set to a triple coincidence with an external NaI(Tl) detector.
This requires a trigger on the S1-signal rather than the much larger S2-signal and causes a clear reduction in the trigger efficiency at low energies, as evidenced by the rate decrease at low energies.

The NR energy spectra for the $^{241}$AmBe and background datasets are shown in figure~\ref{fig:fig5} (right).
The gray points shows the relative contribution of background events within the NR selection.
At low energies, the ER and NR bands overlap, so that the background rate in this region increases.
However, since the majority of NR events occurs at low energy (due to the effect of kinematics and the energy spectrum of the $^{241}$AmBe neutrons), the percentage of background events is low.
At higher energies, the background comes from events that are below the ER band due to incomplete charge collection.
The background rate increases at $\sim$\SI{17}{keV}$_{ee}$ due to the earlier described line feature with significant leakage below the ER band.
We therefore limit the analysis of NR data to $<$\SI{15}{keV}$_{ee}$.
We limit all analysis of electronic recoil data to \SI{70}{keV}$_{ee}$ (roughly corresponding to \num{400}~p.e. cS1 and \num{300000}~p.e. in cS2), as ADC saturation starts appearing above this energy.

\section{Monte Carlo model}\label{sec:mc}
\subsection{Single photoelectron pulse model} \label{sec:spe}
\begin{figure}[h]
\begin{center}
\includegraphics[width=\defaultplotwidth]{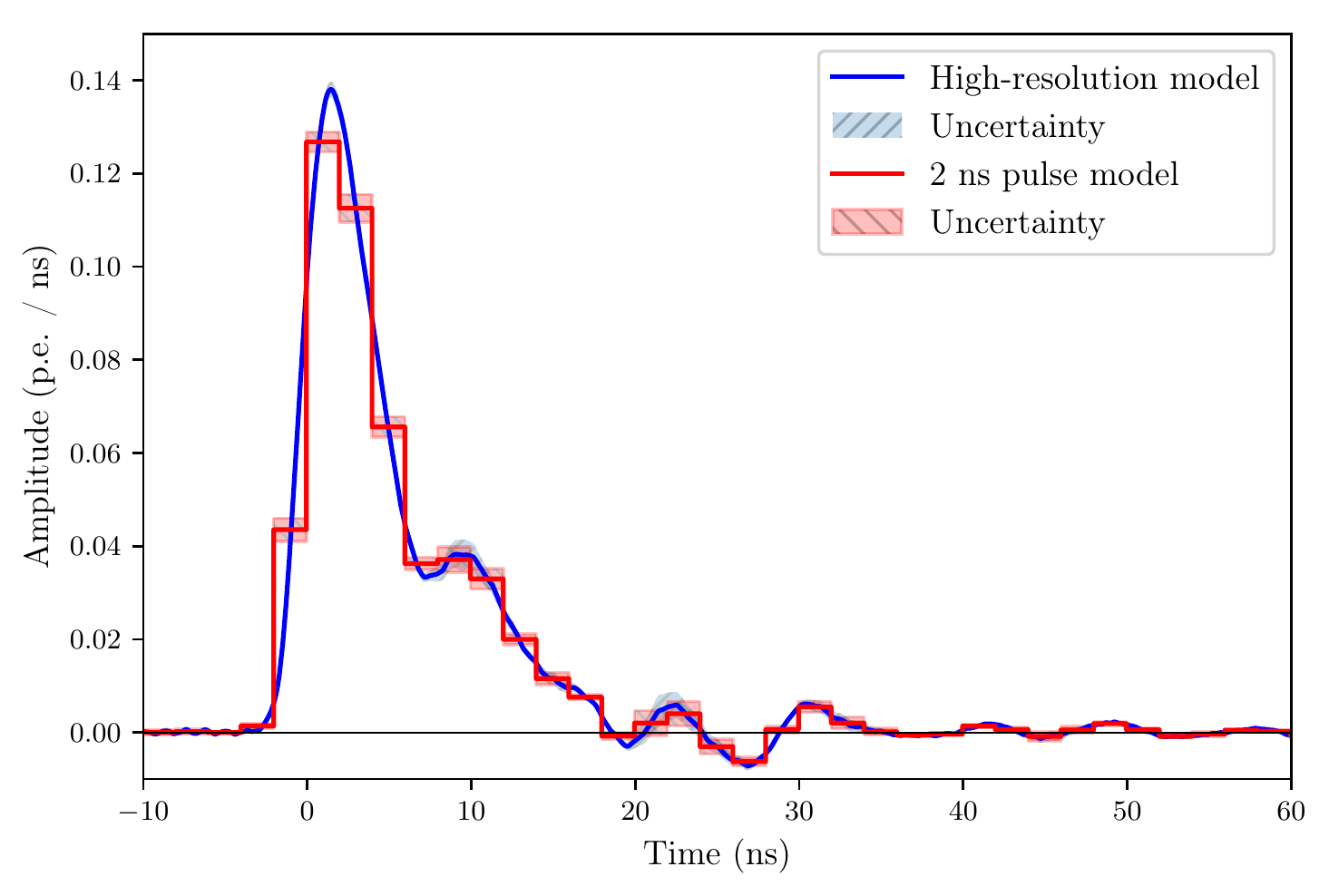}
\caption{Single photoelectron pulse shape model, measured in high resolution (blue, smooth line), and rebinned in \SI{2}{ns} bins (red histogram) as used in the model.
The specific form of the single photoelectron waveform depends on the position of the high-resolution pulse within the \SI{2}{ns} bins.
The uncertainty on the coarse single photoelectron waveform is taken as the difference between the shape obtained from the rebinned high-resolution calibration and the calibration taken in liquid xenon.
This uncertainty is indicated by the bands around the waveforms.
}
\label{fig:fig6}
\end{center}
\end{figure}

Our S1 simulation uses a single photoelectron pulse shape model based on LED calibration measurements using an oscilloscope (Keysight DSO-S 254A) with a time resolution of \SI{0.1}{ns}.
Figure~\ref{fig:fig6} shows the average of single photoelectron waveforms, aligned on the point where \num{10}\% of the area is reached.
We rebin the model to \SI{2}{ns} and apply a \SI{250}{MHz} low-pass frequency filter to correct for the lower time resolution and analog bandwidth of the CAEN V1730D digitizer used in the main measurements.
The resulting pulse model is shown in figure~\ref{fig:fig6}.
We compare the rebinned model to LED calibration data taken at cryogenic temperatures with the V1730D digitizer by computing the average single photoelectron pulse shapes, aligned to the maximum in the sample.
The difference between these is taken as a systematic uncertainty in the main analysis.

The single photoelectron pulse shape has a \SI{5}{ns} width (FWHM) and shows a $\sim$\SI{90}{MHz} oscillation (\emph{ringing}) consistent with standing waves in the $\sim$\SI{2}{m} long cables between the PMTs and the DAQ, presumably due to an impedance mismatch between the PMT bases and the DAQ.
The bottom PMT shows a lower oscillation frequency consistent with its longer cable length.

\subsection{Average pulse shape}
Since our analysis focuses on low-energy recoils, the number of photons in each individual pulse is low.
We therefore use the average pulse shape to fit the measured data to the scintillation model.
The average pulse shape model is constructed in the following way.
A large sample of photon production times with a distribution according to the scintillation model (eq.~\eqref{eqn:2_4} or eq.~\eqref{eqn:2_5}) is generated.
The times are smeared with a normal distribution with mean~\num{0} and a standard deviation which we define as the detector time resolution~$\sigma_{det}$.
The PMT channel for each photon is randomly sampled from a binomial distribution according to the observed light distribution in measured data (with probability of the photon being in the top PMT of \num[separate-uncertainty = true]{0.28 \pm 0.13}).
The area of each single-photon pulse is sampled from the measured area distribution in the single photoelectron gain calibration for the PMT that it was observed by.
The array of photons is then split into S1 clusters, where the areas of the S1s are taken from the area distribution observed in measured data (within the energy selection). 
The S1 pulse shapes are calculated as the sum of the single photon pulse shapes, where the normalization of each single-photon pulse is provided by the sampled area and the time shift is given by the photon arrival time.
We use the high time resolution model described in section~\ref{sec:spe}, rebinned to \SI{0.2}{ns} for ease of computation, as the single-photon pulse model.
The resulting pulse shape is then shifted by a random integer up to \num{10} of \SI{0.2}{ns} bins to simulate the random alignment of the pulse within the digitizer time bins and rebinned to \SI{2}{ns}.
The average pulse shape is computed as the average of all normalized pulse shapes, aligned on the point where \num{10}\% of the area is found.

\subsection{Uncertainties}\label{sec:errs}
We compute the goodness of fit $\chi^2/n_{\rm d.o.f.}$ from
\begin{equation} \label{eqn:4_1}
\frac{\chi^2}{n_{\rm d.o.f.}} = \frac{1}{n_{s} - 1} \sum_{i}^{n_{s}} \frac{(y_{{\rm data}, i} - y_{{\rm model}, i})^2}{\sigma_i^2},
\end{equation}
where~$i$ runs over all \SI{2}{ns} samples in the pulse that are within the fit range, $n_s$ is the number of samples in the waveform (ranging from \num{58} to \num{76} as the fit range is changed, see section~\ref{sec:fit_proc}) and $y$ and $\sigma$ are the average waveform amplitude and its uncertainty.
The average waveform uncertainty we compute is built up of several parts.
The dominant contribution comes from the single photoelectron pulse shape model described in section~\ref{sec:spe}.
We compute the corresponding uncertainty on the average pulse shape by adding the uncertainty on the single photoelectron pulse shape for all photons in quadrature per S1 waveform.
A second contribution of the uncertainty is the statistical uncertainty due to the finite number of S1 waveforms in the data sample.
We compute this by resampling the data waveforms \num{250} times and recomputing the average waveform, where we define the uncertainty as the per-sample standard deviation.
A third source of uncertainty comes from the distribution of photons across both PMTs, which varies stochastically in the data with changing light detection efficiencies at different interaction positions. 
We therefore recompute the average waveform with light fraction in the top PMT of \num[separate-uncertainty = true]{0.28 \pm 0.13} (corresponding to $1 \sigma$) and take the variation on the average waveform as a systematic uncertainty.
Finally, we add a constant uncertainty of $10^{-4}$ in units of fraction of amplitude, corresponding to approximately \num{0.1} - \num{0.2}\% of the maximum pulse amplitude, to account for any unmodeled uncertainties (such as the influence from randomly distributed noise hits or dark counts in the waveform).
This uncertainty is subdominant in all parts of the waveform, but stabilizes the fit if all the other uncertainties fluctuate downward (causing very large values of $\chi^2/n_{\rm d.o.f.}$ because of the $1/\sigma^2$ dependence in eq.~\eqref{eqn:4_1}).
All these uncertainties are added in quadrature to yield the per-sample uncertainty on the average waveform.

\subsection{Fit procedure}\label{sec:fit_proc}
\begin{figure}[h]
\begin{center}
\includegraphics[width=0.5 \linewidth]{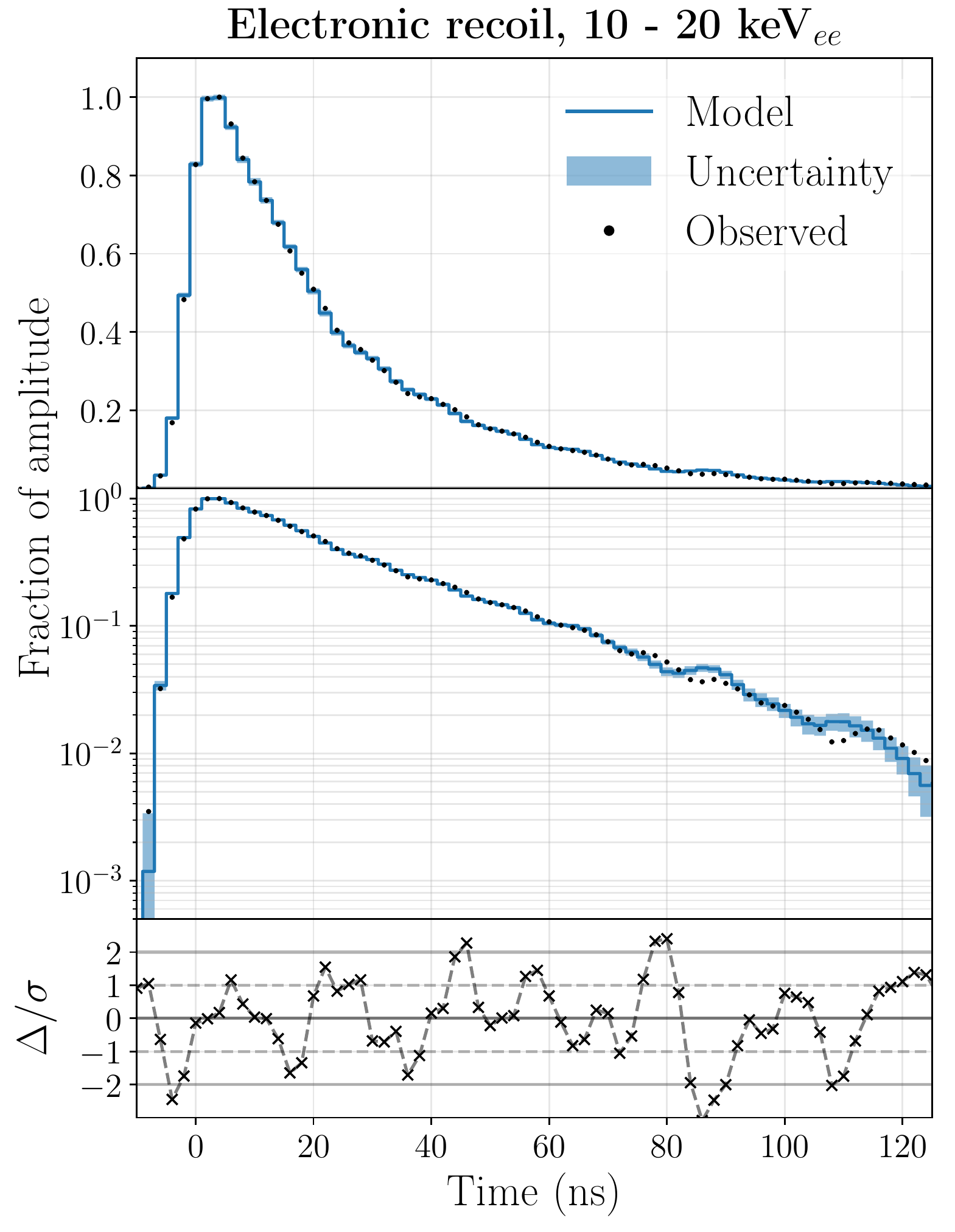}
\caption{
Example of an average waveform shape for electronic recoil data between~\num{10} and \SI{20}{keV}$_{ee}$, taken at \SI{0.5}{kV/cm}, shown on a linear scale (top panel) and logarithmic scale (middle panel),
Measured data is indicated by the black dots, while the blue line and the shaded blue band show the best-fit simulated pulse shape and its uncertainty.
The bottom panel shows the residuals, normalized by the uncertainty.
An oscillatory behavior is seen for both the data and the model, coming from the ringing observed in the single photoelectron waveforms (see figure~\ref{fig:fig6}).
The value of $\chi^2/{n_{\rm d.o.f.}}$ for this fit is 1.44.}
\label{fig:fig7}
\end{center}
\end{figure}

We produce the average pulse shape for various parameter combinations of both scintillation models (eq.~\eqref{eqn:2_4} or eq.~\eqref{eqn:2_5}) on a parameter grid.
For the exponential fit, the fit parameters are $f_s$, $\tau_s^{\rm eff}$, $\tau_t^{\rm eff}$ and the detector resolution $\sigma_{det}$ and the grid spacing is \num{0.005}, \SI{0.25}{ns}, \SI{0.25}{ns} and \SI{0.125}{ns}, respectively.
For the recombination model, some of the parameters are fixed (see section~\ref{sec:fit_recomb}), leaving $T_R$, $\eta$, $f_R$ and $f_s^R$ with a grid spacing of \SI{1}{ns}, \num{0.05}, \num{0.025} and \num{0.05}, respectively.
The best-fit parameters are taken as the point where the minimum value of $\chi^2/n_{\rm d.o.f.}$ is reached.
Figure~\ref{fig:fig7} shows an example for one of the waveform fits.

We validate the fit procedure by generating a sample of waveforms with known parameter values and then fitting this in the same way as done with measured waveforms.
This is done for each energy bin in each dataset with the same number of S1 waveforms as observed in measured data.
In this way, we simultaneously evaluate the statistical uncertainty, which we define as the standard deviation of the parameter around the fit point.

There are also systematic uncertainties coming from the freedom in the definition of the fit range and possible errors in the estimate of the uncertainty of the average waveform.
We estimate the effect of the fit range by varying the left and right boundary within \SIrange{-14}{-6}{ns} and \SIrange{110}{140}{ns} (with respect to the \num{10}\% alignment point).
To quantify the uncertainty of the fit parameters coming from the uncertainty estimate on the average waveform, we vary the constant uncertainty described in section~\ref{sec:errs} from \num{0} to \num{2e-4}.
These three parameters are all varied simultaneously and the best-fit point is recomputed.
We take the uncertainty on the best-fit point as the standard deviation of the parameters.

\begin{figure}
\begin{center}
\includegraphics[width=\wideplotwidth]{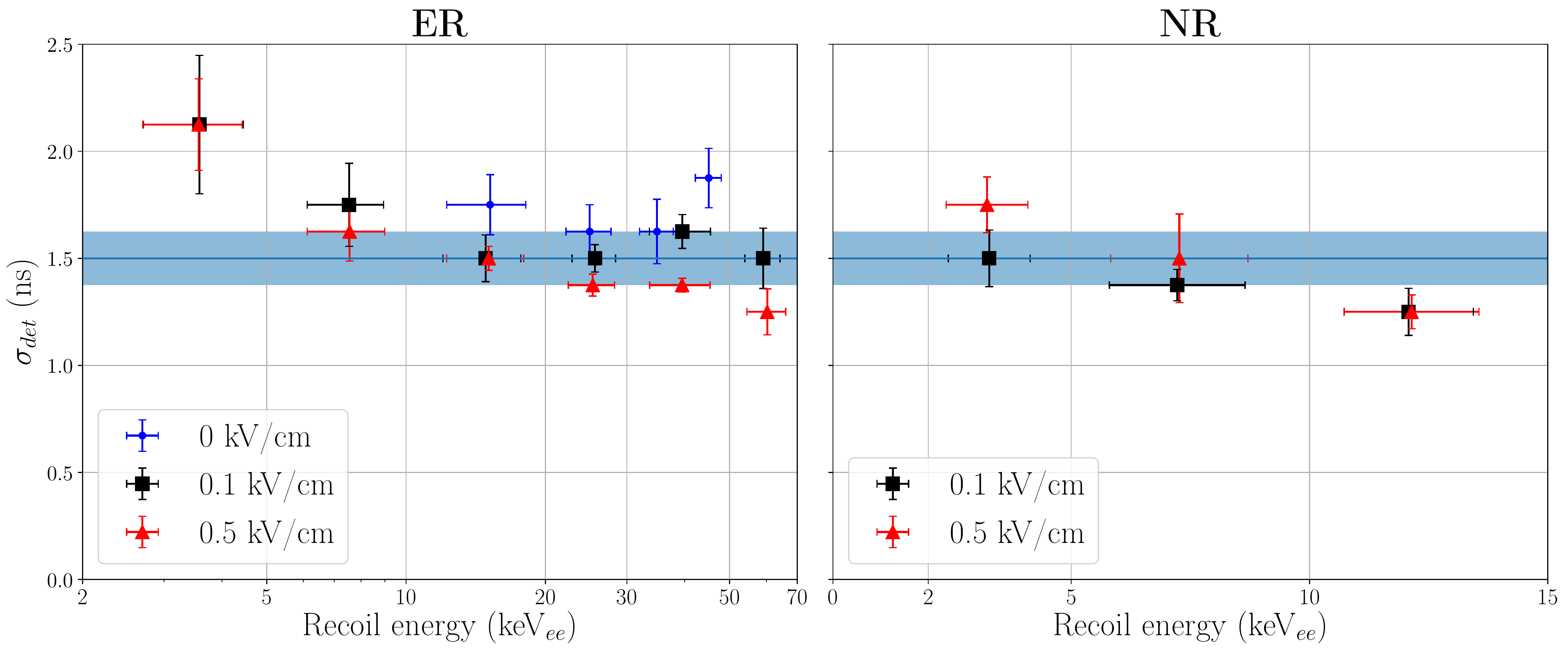}
\caption{The detector resolution~$\sigma_{det}$ for all ER data (left panel) and NR data (right panel).
    The values shown here were obtained by applying an exponential fit (eq.~\ref{eqn:2_5}) with a Gaussian smearing of the photon times with standard deviation~$\sigma_{det}$.
    As the detector resolution is a detector quantity, we fix the value of this parameter to~\SI{1.5}{ns} for all fits.
    We estimate the uncertainty to be~\SI{0.1}{ns}, and vary~$\sigma_{det}$ with one step in the parameter grid ($\pm$ \SI{0.125}{ns}, as indicated by the blue band) to evaluate the effect on the other fit parameters.
}
\label{fig:fig8}
\end{center}
\end{figure}

The detector resolution~$\sigma_{det}$ is a quantity that depends on the setup, hence it should be a constant parameter independent of the measurement.
Nevertheless, we allowed this quantity to vary for all exponential fits.
Figure~\ref{fig:fig8} shows a compilation of the values of~$\sigma_{det}$.
From this, we estimate the detector resolution is \SI[separate-uncertainty=true]{1.5 \pm 0.1}{ns}.
This is in agreement with the expected resolution given the PMT transit time spread of $\sim$\SI{0.75}{ns} \cite{hamamatsu} and the influence of photon travel time due to reflections.
As the length of the TPC is approximately~\SI{10}{cm}, we expect typical photon path differences to be of the same order, giving an additional spread of~$\sim$\SI{0.5}{ns}.\footnote{Note the effective speed of light in liquid xenon is lower due to the index of refraction of \num[separate-uncertainty=true]{1.69 \pm 0.02} \cite{solovov2004}.}
For the rest of the analysis, the detector resolution is fixed at~\SI{1.5}{ns}.
We determine any uncertainty on the other fit parameters by varying $\sigma_{det}$ by \SI{0.125}{ns}, corresponding to one step in the parameter grid spacing of~$\sigma_{det}$.

\section{Results}

\subsection{Double-exponential fit} \label{sec:fit_exp}
\begin{figure}
\begin{center}
\includegraphics[width=\wideplotwidth]{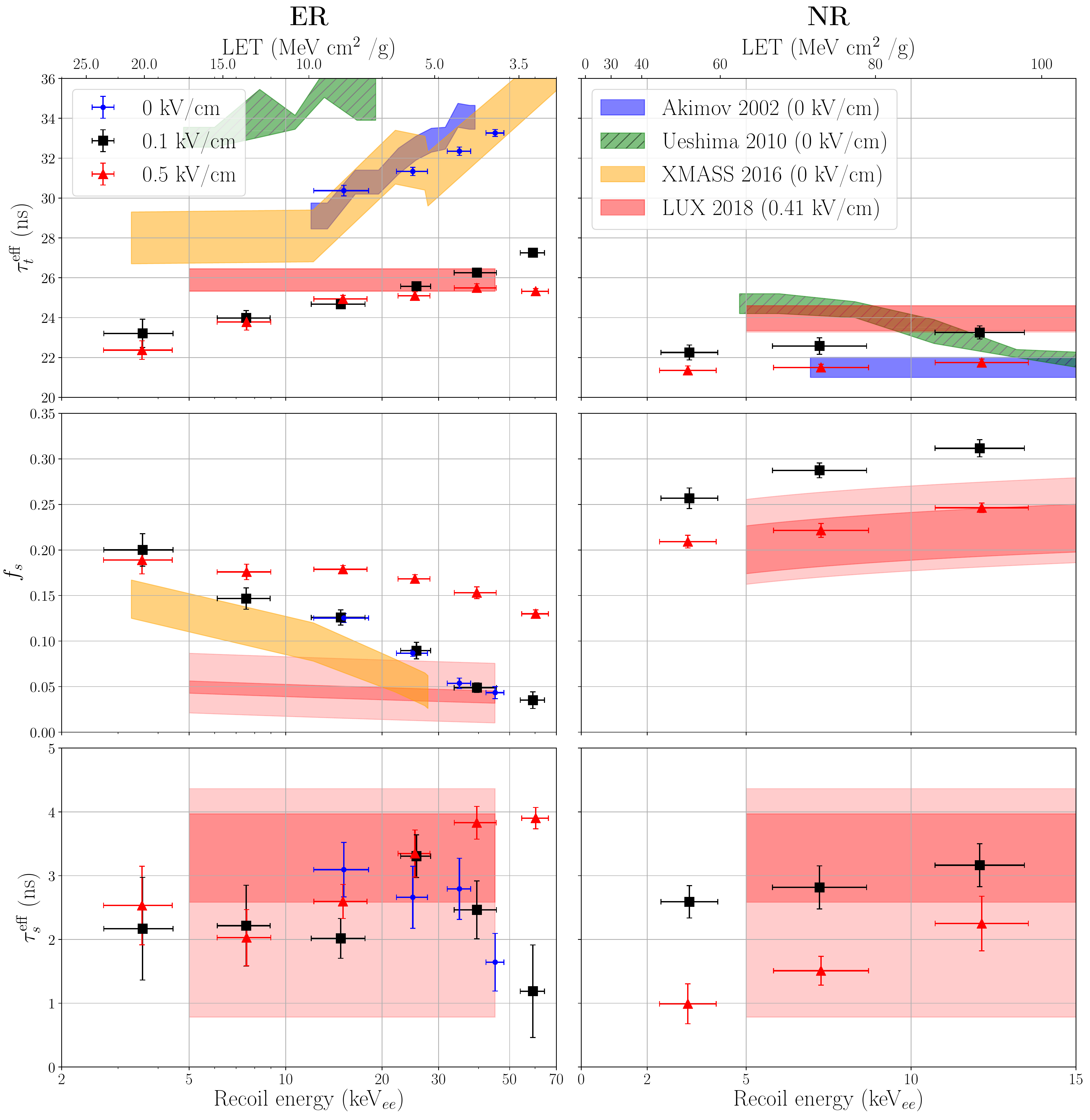}
\caption{The best-fit parameters using a double-exponential model (eq.~\eqref{eqn:2_5}) for electronic recoils (left row) and nuclear recoils (right row) as a function of recoil energy.
    The electronic linear energy transfer (LET) is indicated on the top scales. 
    The black squares and the red triangles show the data at low field (\SI{0.1}{kV/cm}) and high field (\SI{0.5}{kV/cm}), the blue circles are for zero field data (ER only). 
    The error bars indicate the statistical and systematic uncertainty, described in section~\ref{sec:fit_proc}, added in quadrature.
    The triplet lifetimes as measured by Akimov et al. \cite{akimov2002}, Ueshima \cite{ueshima_phd} and the triplet lifetimes and singlet fractions measured by XMASS \cite{xmass2016} are indicated by the blue, hatched green and orange bands, respectively.
    Note that these measurements were performed at zero field, and fit using a single exponential distribution for Akimov et al. and Ueshima.
    The LUX measurements (\SI{0.41}{kV/cm}) are indicated by the red bands, with the light shaded band indicating the systematic uncertainty due to the optical simulation  \cite{luxpsd}.
}
\label{fig:fig9}
\end{center}
\end{figure}

The data taken consists of five measurement series; two for the two different fields for NR data and three for the ER data that includes data taken at zero field.
The data is binned in energy as obtained from the combined energy scale for all the points with nonzero field and from the cS1 in the case of zero field.
The value and the error of the energy for each bin is set to the mean and standard deviation of the energy in the bin.
Figure~\ref{fig:fig9} shows the fit values for a double-exponential fit on ER (left panels) and NR (right panels) data, at \SI{0.5}{kV/cm} and \SI{0.1}{kV/cm} and including zero field for ER.
The bottom scales show the measured recoil energy, while the top scales show the corresponding electronic LET (obtained using the ESTAR \cite{ESTAR} and the SRIM \cite{SRIM} models for ER and NR, respectively), as given in \cite{aprile2006}. The recoil energy for nuclear recoils was calculated using the quenching factor parametrized in \cite{nest2}.
The shaded bands in the figures show results from  Akimov et al. (blue band) \cite{akimov2002} and Ueshima (green hatched band) \cite{ueshima_phd}, both using data at zero field and a single exponential fit, measurements from XMASS (orange band) \cite{xmass2016} at zero field and results from LUX (red bands) \cite{luxpsd}, where the light shaded bands indicate the systematic error due to uncertainties in the optical model.
The field for the LUX data was \SI{0.41}{kV/cm}, comparable to \SI{0.5}{kV/cm} as used for our high-field measurements.

The top two panels show the effective triplet lifetime~$\tau_t^{\rm eff}$ as a function of energy for ER and NR.
For all datasets, the apparent lifetime increases with energy, though the increase in the NR datasets is mild and could at least partly be explained by the $\mathscr{O}(10 \%)$ ER contamination described in section~\ref{sec:data}. 
The values found converge to \SI[separate-uncertainty=true]{22 \pm 1}{ns} for low energy nuclear recoils at both field strengths, in agreement with (zero-field) measurements of low-energy nuclear recoils reported by Akimov et al., but disfavoring values found by Ueshima and LUX.
For ER, there is an increase of the effective triplet lifetime that becomes stronger for low field strengths and high energy due to the non-negligible recombination time.
This is in disagreement with the analysis followed by LUX, where a recombination time $<$\SI{0.7}{ns} was assumed based on an extrapolation of an empirical formula to low energies \cite{mock2014}.
In contrast, at a field of \SI{0.5}{kV/cm}, we find an increase from roughly \num{22} to \SI{25}{ns}, giving a recombination time of at least $\sim$\SI{3}{ns} at \SI{50}{keV}$_{ee}$.
The high-energy limit of \SI{25}{ns} seems to correspond reasonably well with the value of \SI[separate-uncertainty=true]{27 \pm 1}{ns}, which was found at a higher field strength (\SI{4}{kV/cm}) and a higher energy (from a $^{207}$Bi source, typically \si{MeV} electrons).
For zero field, the apparent triplet lifetime component is significantly longer and seems to increase to beyond the observed maximum of \SI[separate-uncertainty=true]{33 \pm 1}{ns}. 
It is therefore unclear if the high-energy limit found here favor the values found in literature of~\SI[separate-uncertainty = true]{34 \pm 2}{ns} \cite{kubota1978} and \SI[separate-uncertainty = true]{33 \pm 1}{ns} \cite{keto} or $\sim$\SI{45}{ns} \cite{kubota}.
We find good agreement with the values found by Akimov et al. \cite{akimov2002} and XMASS \cite{xmass2016}, but significant disagreement with measurements from Ueshima \cite{ueshima_phd}.

The second row in figure~\ref{fig:fig9} shows the singlet fraction as a function of energy.
The ER and NR datasets are showing opposing trends; for higher energies, the singlet fraction rises for NR and decreases for ER events.
This suggests that $f_s$ is correlated with the LET (shown on the top scales), which decreases as a function of recoil energy for ER but increases for NR.
This is in agreement with previous measurements that found a higher value of $f_s$ for particle species that have a higher ionization density.
For high fields, the influence is reduced, pointing to the recombination process as a physical origin of this difference.
We find reasonable agreement with the zero-field measurements from XMASS.
The LUX measurements showed compatibility with a linear or an exponential model.
We find better agreement with the exponential trend and show these bands for comparison.
While the NR data confirms LUX's result, the values of $f_s$ found for ER show a strong disagreement; the values found by LUX of $\sim$\num{0.04} are much lower than the high field points around \num{0.16} for this energy range and electric field.

The bottom row in figure~\ref{fig:fig9} shows the apparent singlet lifetime~$\tau_s^{\rm eff}$.
For this parameter, the fit uncertainties are much larger than for the first two parameters.
This is partly due to a correlation between $\tau_s^{\rm eff}$ and $\sigma_{det}$, which can be understood from the waveform model: both have the net effect of smearing out the strongly peaked part of the average waveform.
In addition, the short timescales with respect to the digitizer time resolution makes the fit value depend strongly on only a few samples around the maximum, making the fit sensitive to mismodeling of the waveform shape.
Given these caveats, we find decent agreement with the singlet lifetimes in the range of \SIrange{2.2}{4.3}{ns} found in literature \cite{hitachi, luxpsd}.
 
\subsection{Recombination model} \label{sec:fit_recomb}
Rather than using the simple exponential model, we can use the full recombination model introduced in section~\ref{sec:recomb_model}.
This is a seven-parameter model, given by eq.~\eqref{eqn:2_4}, so that an unconstrained fit would be challenging due to correlation between the parameters.
However, since the scintillation model is split into a part that is due to direct excitation (dependent on $\tau_s$, $\tau_t$ and $f_s$) and a part due to recombination luminescence (dependent on $T_R$, $f_s^R$ and $\eta$), we can use the parameters of $\tau_s$, $\tau_t$ and $f_s$ that are found for the cases where the influence of recombination is negligible, such as for nuclear recoils at high field.
We therefore fix the values of $\tau_s$, $\tau_t$ and $f_s$ to \SI{1.5}{ns}, \SI{21.5}{ns} and \num{0.23}, respectively.
The detector resolution~$\sigma_{det}$ is kept fixed at \SI{1.5}{ns}.
This leaves four parameters: the recombination time~$T_R$, the electron escape probability $\eta$, the fraction of photons due to recombination~$f_{R}$ and the fraction of recombination photons coming from the singlet state~$f_s^R$.
The resulting four-parameter fit was performed for the zero field ER data.

\begin{figure*}
    \makebox[\linewidth]{
        \includegraphics[width=\linewidth]{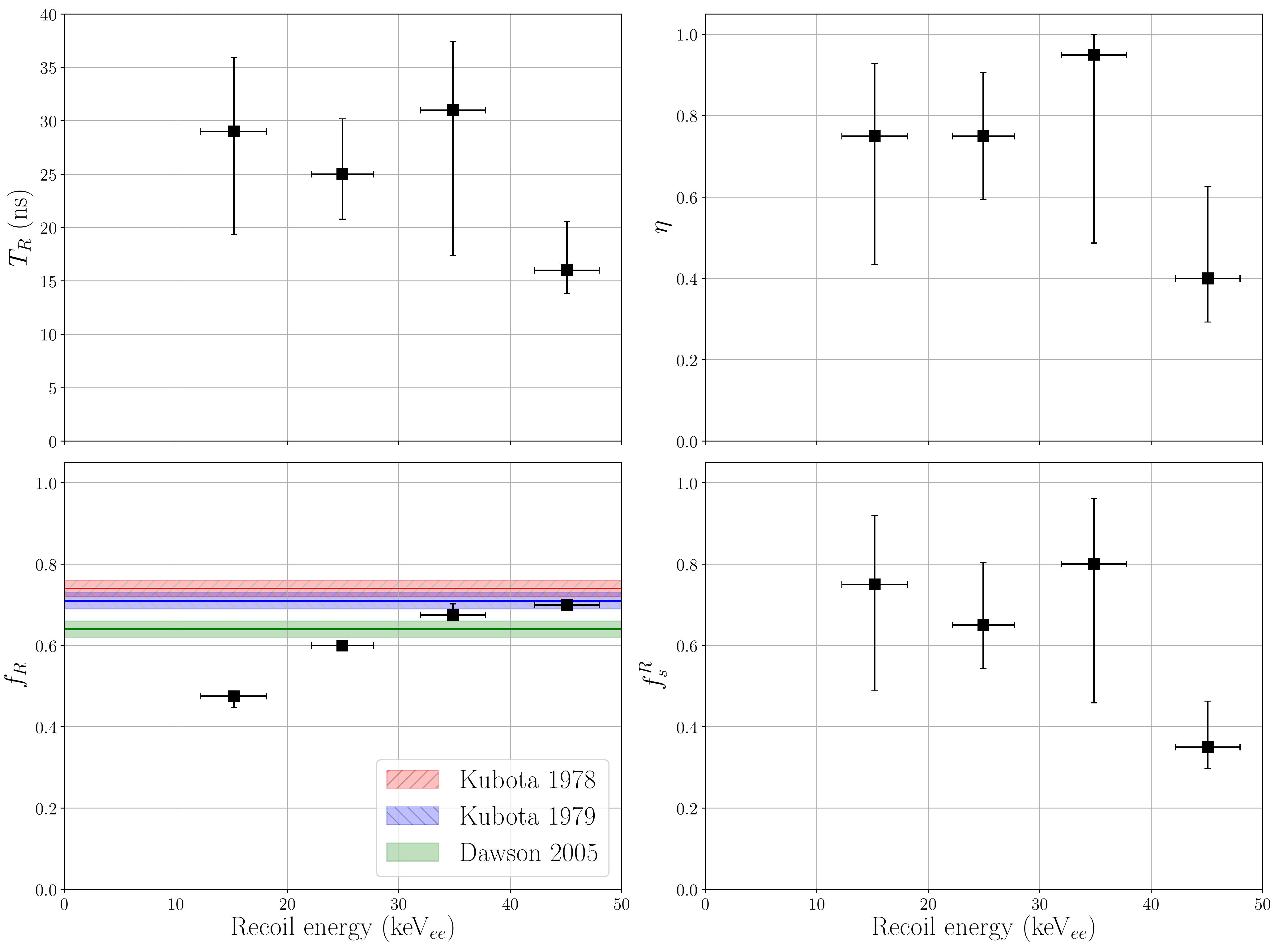}
    }
    \caption{The best-fit parameters for the model including recombination introduced in section~\ref{sec:recomb_model}, described by eq.~\eqref{eqn:2_4}. 
    The fit is applied to the ER data taken at zero field.
    The panels show the fit parameters for the recombination time~$T_R$, the electron escape probability~$\eta$, the recombination fraction~$f_R$ and the recombination singlet fraction~$f_s^R$.
    The parameters $\tau_t$, $\tau_s$, $f_s$ and the detector resolution~$\sigma_{det}$ are fixed at \SI{21.5}{ns}, \SI{1.5}{ns}, \num{0.23} and \SI{1.5}{ns}, respectively.
    The recombination fraction~$f_R$ increases with energy up to values consistent with measurements performed at high energies, shown by the colored bands \cite{kubota1978, kubota, dawson2005}.}
    \label{fig:fig10}
\end{figure*}

Figure~\ref{fig:fig10} shows the resulting best-fit values.
All the uncertainties are determined in the same way as described in section~\ref{sec:fit_exp}.
The uncertainties of the recombination time $T_R$, the escape fraction $\eta$ and the singlet recombination fraction $f_s^R$ are substantial due to the observed correlation of these parameters.
This makes it difficult to draw quantitative conclusions for these parameters. 
The recombination time~$T_R$ and the recombination singlet fraction~$f_s^R$ at high energies seems to be consistent with the values found by Kubota et al. of \SI[separate-uncertainty = true]{15 \pm 2}{ns} and \num[separate-uncertainty = true]{0.44 \pm 0.11}~\cite{kubota}, though it should be noted that this was fit for high-energy electrons and using a different value for $\eta$.
Rather than the values $\eta < 0.05$ found for high-energy electronic recoils, we find that the escape probability is significantly higher at the energies studied here ($>$\num{0.4} for $E<$\SI{40}{keV}$_{ee}$).

The most notable result from the fits, is the fraction of scintillation light coming from recombination~$f_R$. 
This value may be compared to the measurements of luminescence quenching with a sufficiently high electric field so that all electrons are pulled away from the interaction site, which indicate a value of around~\num{0.7} as indicated by the colored bands in figure~\ref{fig:fig10} \cite{kubota1978, kubota, dawson2005}.
At high energies, where these measurements were performed, the results obtained here show excellent agreement; the recombination fraction appears to asymptotically approach this value.
It should be noted that the method described here is completely independent of the measurements that rely on extracting the electrons from the interaction site: the same result is found without applying any electric field.

In general, if the results of Kubota et al. are compared to the low-energy equivalent presented here, the conclusion can be drawn that the recombination  efficiency decreases at low energies.
This is evidenced by the decrease in $f_R$ with respect to the constant value of $0.7$ and the increase in $\eta$ towards low energies.
This conclusion is consistent with recent measurements that indicate a decreasing scintillation yield and an increasing charge yield for low-energy electronic recoils~\cite{goetzke2017}.
Furthermore, we note that the apparent decrease of $f_s$ at high energy as observed in the double-exponential fit (section~\ref{sec:fit_exp}) is modeled without including superelastic collisions.
Instead, this is accounted for in the following way.
At high energy, $f_R$ increases, so that a larger fraction of the luminescence comes from recombination.
Since the recombination timescale is long with respect to the singlet lifetime, most of the recombination luminescence is fitted by the triplet component in the double-exponential model, reducing the value of $f_s$ where recombination is dominant.

In this section, we have shown that a relatively simple recombination model can be used to explain the pulse shape difference observed at zero field.
However, it should be noted that the true physical process can be more complex and may not be fully captured by the model.
In particular, the ad-hoc parameter~$\eta$ could be replaced by a more detailed description of the dynamical behavior of electrons.
It is furthermore possible that the quenching processes and superelastic collisions described in section~\ref{sec:model} give a non-negligible influence on the pulse shape.
Finally, the results here depend on assumed values of $\tau_s$, $\tau_t$ and $f_s$, which were estimated from figure~\ref{fig:fig9}.

\section{Application to pulse shape discrimination}
As liquid xenon TPCs are currently leading the field of direct dark matter discrimination, there has been considerable interest in using pulse shape discrimination as a method of reducing the electronic recoil background in large-scale TPCs \cite{ueshima2011,kwong2010,  luxpsd}.
The theoretical performance of PSD in large-scale dual-phase detectors was previously studied by Kwong et al. \cite{kwong2010}, who found that using PSD was not performing well enough to be productive in large-scale TPCs looking for dark matter elastic recoils.
This was concluded based on the comparison of the PSD performance with the regular S2/S1 discrimination.
In the LUX detector, a combined discriminator decreased the ER contamination from \SI[separate-uncertainty=true]{0.4 \pm 0.1}{\percent} to \SI[separate-uncertainty=true]{0.3 \pm 0.1}{\percent} at a fixed NR acceptance of \SI{50}{\percent} \cite{luxpsd}.
In this section, we explore the theoretical performance of a combined discrimination in XENON1T.
We use simulated data from the Laidbax code \cite{laidbax} combined with Blueice \cite{blueice} to compute the likelihood function.

Currently, the method used by XENON1T to determine WIMP-nucleon interaction cross section limits uses the likelihood ratio computed in two dimensions, S1 and S2 \cite{xe1t}.
This may be extended to a three-dimensional parameter space, using a PSD parameter as the third parameter.
Given the description of the pulse shape parameters at low energies shown in section~\ref{sec:fit_exp}, we can calculate the performance of such a combined discrimination in an idealized experiment.

We produce Monte Carlo samples of a flat electronic recoil background component and that of a WIMP recoil.
For each Monte Carlo generated event, we sample photon production times from the distribution described by eq.~\eqref{eqn:2_5} with $\tau_s^{\rm eff}$, $\tau_t^{\rm eff}$ and $f_s$ taken from an interpolation of the best-fit values displayed in figure~\ref{fig:fig9}.
As the drift field of XENON1T is close to~\SI{0.1}{kV/cm}, we take the values found for the low drift field.
The times are then smeared with an adjustable detector resolution~$\sigma_{det}$.
The PSD parameter used is the logarithm of the likelihood ratio that is calculated from the photon probability density functions $f_{ER}(t)$ for electronic recoils and $f_{NR}(t)$ for nuclear recoils.
The likelihood ratio~$\mathscr{L}$ is computed as,
\begin{equation} \label{eqn:6_1}
\mathscr{L} = \frac{\prod_{t \in T} f_{NR}(t; E, \sigma)}{\prod_{t \in T} f_{ER}(t; E, \sigma)},
\end{equation}
where $T$ is a set of photon detection times and where the energy $E$ is the reconstructed ER equivalent energy (the dependence on~$E$ is through the dependence of the pulse shape parameters).
The numerator and the denominator give the likelihood of the times in~$T$ to be drawn from the distributions for nuclear recoils and electronic recoils, respectively, so that a high value is classified as an `NR-like' event, while a low value is likely to be an ER event.
We maximize both likelihoods with respect to the start time of the pulse.
The discrimination efficiency depends on the scintillation and ionization gains~$g_1$ and $g_2$; the values taken here are \num{0.144} and \num{11.5}, respectively, as found for the first results of XENON1T.
We simulated WIMP spectra for \SI{20}{GeV}, \SI{50}{GeV} and \SI{500}{GeV} and a time resolution of \num{10}, \num{5} and \SI{2}{ns}, roughly representing the current time resolution, an improved time resolution and a significantly improved time resolution.
It should be noted that the estimate of \SI{10}{ns} is optimistic, given the typical PMT transit time spread of \SI{9}{ns}~\cite{xenon1tpmts} and the digitizer resolution of~\SI{10}{ns}~\cite{xenon1tdesign}.
For all the simulated datasets, over two million events were simulated.

We evaluate the ER and WIMP separation efficiency by constructing the receiver operating characteristic (ROC) curve, which gives the WIMP acceptance fraction as a function of ER acceptance.
A three-dimensional histogram of the parameters S1, S2 and $\mathscr{L}$ is constructed for the WIMP signal and the ER background data.
For each bin, the signal-to-background ratio is evaluated based on the number of WIMP and background events in the bin.
Given the minimum signal-to-background ratio allowed, the ER and WIMP acceptance may be calculated as the fraction of all events in the bins with a higher signal-to-background ratio.
The ROC curves for S2/S1 or PSD only are evaluated similarly, with two- and one-dimensional histograms.
The allowed range for all events is \numrange{3}{70}~\!p.e. for S1 and \numrange{50}{8000}~\!p.e. for S2, corresponding to the analysis range used for XENON1T's first results.
All acceptances are expressed as a fraction of events after passing the S1 and S2 selection criteria.

\begin{figure}
\begin{center}
\includegraphics[width=0.9 \linewidth]{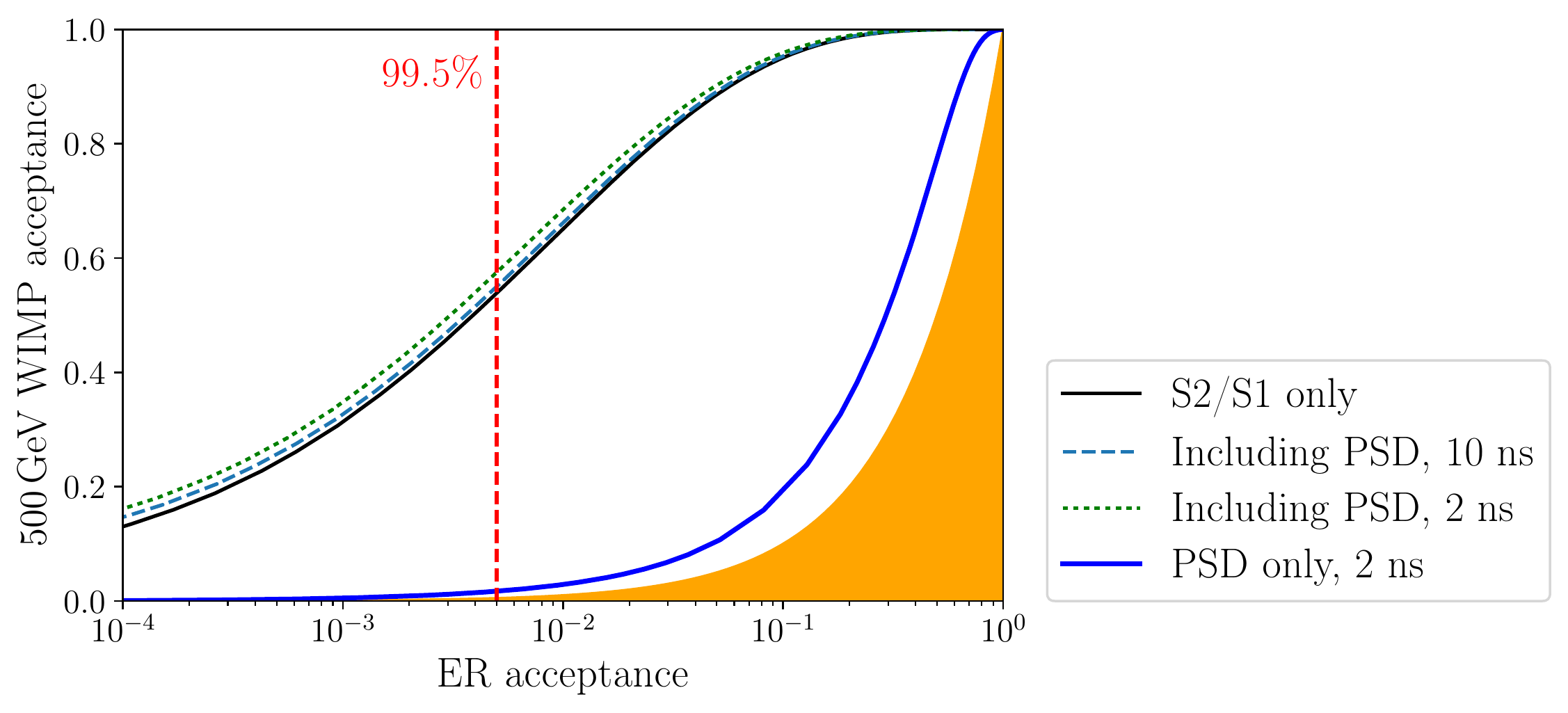}
\caption{Simulated acceptances of events from a \SI{500}{GeV} WIMP recoil as a function of the acceptance of a flat ER background for four different discrimination methods.
The solid black line shows the results from using only the information of the S1 and the S2 size, while the dashed blue and dotted green lines are the performance when this method is combined with PSD, given a detector timing resolution of \num{10} or \SI{2}{ns}.
The thick solid blue line shows the performance using PSD alone, while the yellow shaded area is bounded by the line~$y=x$ that gives the acceptance of a random cut.
At an ER rejection of \num{99.5}\%, the WIMP acceptance rises by \num{6.8}\% if PSD is included, given a time resolution of \SI{2}{ns}.
For lower WIMP masses, the PSD performance decreases even more.
}
\label{fig:fig11}
\end{center}
\end{figure}

Figure~\ref{fig:fig11} shows four ROC curves for a~\SI{500}{GeV} WIMP.
The solid black line shows the discrimination achieved by using only the position of the events in (S1,S2)-space, neglecting all pulse time information.
The dashed blue and dotted green lines show the improvement that can be achieved by including the pulse time information.
As these lines barely improve over the baseline scenario, the improvement by including PSD is marginal.
At an ER background rejection of \num{99.5}\% (indicated by the red dashed line), which is a typical value used for  liquid-xenon based dark matter searches, including PSD increases the WIMP acceptance from \num{54.0}\% to \num{57.7}\% even for a time resolution of \SI{2}{ns}, giving a relative exposure increase of \num{6.8}\%.
At lower WIMP masses, this drops to even lower values (3.9\% increase at \SI{50}{GeV}, 0.6\% at \SI{20}{GeV}), as may be expected due to the decreased performance of PSD at the low energies that are more relevant for low WIMP masses.
The thick blue line shows the ROC curve for PSD only.
The shaded yellow portion of the graph is bounded by the line~$y=x$, which would be the ROC curve from a fully random cut on the data.

The LUX collaboration has used the pulse shape information to decrease the ER background leakage in the LUX detector from \SI[separate-uncertainty = true]{0.4 \pm 0.1}{\%} using only the S2/S1 ratio to \SI[separate-uncertainty = true]{0.3 \pm 0.1}{\%} using a combination of S2/S1 and PSD \cite{luxpsd}.
This number is the average leakage over the range of \num{10} to \num{50} p.e.\! at a fixed NR acceptance of $\sim$\SI{50}{\%}.
Although a full simulation of the LUX detector requires a detailed understanding of the LUX detector that goes beyond the scope of this work, we determine the  background leakage measured in the same way in our simulation, setting the time resolution to LUX's value of \SI{3.8}{ns}.
This gives a background decrease from \SI[separate-uncertainty = true]{0.47}{\%} using S2/S1 only to \SI[separate-uncertainty = true]{0.40}{\%} including PSD, in reasonable agreement with the measured values.

It should be noted that the performance of PSD as given in figure~\ref{fig:fig11} depends on several assumptions.
Firstly, the scintillation yield was set to the XENON1T value. 
If the light yield increases, there are more photons available to increase the statistical power of the discrimination in eq.~\eqref{eqn:6_1}.
This will also ameliorate the S2/S1 discrimination.
Secondly, the pulse shape parameters depend on the applied field.
If the drift field is lower, the ER pulse shape broadens due to the longer effective triplet lifetime and lower singlet fraction (see figure~\ref{fig:fig9}), while the NR pulse shape appears to shorten (as seen by the increase in $f_s$ for the lower field data), so that the pulse shapes become more distinct.
Given the recent trend of large-scale detectors to operate at a lower drift field, this may become more relevant for next generation detectors.
Finally, the simulation assumes an ER background in (S1,S2)-space, neglecting anomalous backgrounds formed by accidental coincidence of S1- and S2-signals, wall events that have reduced S2 sizes, or unknown causes.
These backgrounds may escape the S2/S1 discrimination, so that PSD could be used as an independent method to determine their origin or suppress them.

\section{Conclusions}
We have measured the scintillation pulse shape for electronic and nuclear recoils in liquid xenon down to low energies at different field strengths.
By using a Monte Carlo based pulse shape description, we have probed the effective singlet and triplet lifetimes and the fraction of light coming from the singlet state.
The high-energy limits of the results are compatible with previous measurements taken at higher energies.
The measurements for electronic recoils are in tension with the measurements from LUX, as we found significantly higher values for the singlet fraction and a non-negligible recombination timescale for electronic recoils.
A model of the time dependence for the recombination luminescence was taken from Kubota et al.~\cite{kubota} and applied to the electronic recoil data taken at zero field.
Although correlation between the fit parameters makes it difficult to draw solid conclusions, we show that a good description of the pulse shape may be found by using a model that accounts for the recombination luminescence time dependence.
Moreover, the fitted fraction of light from recombination shows excellent agreement with independent measurements.
Given the measured pulse shapes, the theoretical WIMP/background discrimination performance of a large-scale dual-phase xenon detector was computed for a method combining the S2/S1 method and PSD in an optimal way.
It is shown that even with a very fine time resolution and a relatively high-mass WIMP, the exposure increase is marginal, so that PSD will likely not be used as a main method for discriminating elastic WIMP recoil interactions.

\appendix

\section{Recombination model} \label{app:rec}
In this model, we consider a region with volume~$V$ where there is a uniform density of free electrons~$n_e$ and ions~$n_{ions}$ caused by the initial ionizing particle.
Both of these densities are time-dependent, as electrons and ions recombine to form excited states.
 We describe the number density of the recombination electrons $n_e$ by the differential equation
\begin{equation}
\frac{d n_e}{d t} = \frac{d n_{ions}}{d t} = - \alpha  n_e(t) n_{ions}(t).
\end{equation}
If the free electrons and ions are completely confined to the volume~$V$, the initial values of $n_e$ and $n_{ions}$ as well as their derivatives are equal.
We can then define $n(t) \equiv n_e(t) = n_{ions}(t)$ and $n(0) = n_0$, giving the solution
\begin{equation}
n(t) = \frac{n_0}{1+ \frac{t}{T_R}},
\end{equation}
where $T_R$ is the recombination time constant given by $(n_0 \alpha)^{-1}$. 
This case corresponds to full recombination, as for $t \rightarrow \infty$ all electrons and ions have recombined ($n \rightarrow 0$).
However, since the electrons are initially energetic with respect to the energy associated with the Coulomb binding energy, there is a finite probability~$\eta$ that these electrons escape the region~$V$ where recombination occurs shortly after the primary interaction.
In this case, the boundary conditions change to $n_{ions}(0) = n_0$ and $n_{e}(0) = (1 - \eta)n_0$,  and the equation has to be solved numerically.
If we assume that the recombination electrons and ions form an excited molecular state in either the triplet or singlet state, the corresponding number of excited states $N$ with lifetime~$\tau$ is given by
\begin{eqnarray} \label{eqn:A_3}
\frac{dN}{dt} &=& - \frac{N(t)}{\tau} +  - V \frac{d n_e}{dt}  \nonumber \\
&=&- \frac{N(t)}{\tau} + V \alpha n_e(t) n_{ions}(t),
\end{eqnarray}
with the two terms on the right hand side corresponding to the decay and the production of the excited states.
This equation has the solution:
\begin{equation}
N(t) = V \alpha \exp{\left( \frac{-t}{\tau} \right) }  \int_0^t  n_e(t')n_{ions}(t')  \exp{\left(\frac{t'}{\tau}\right)}\ dt'.
\end{equation}
The luminescence time dependence~$I(t)$ is equal to $\frac{1}{\tau} N(t)$, which gives:
\begin{equation}
I(t) = \frac{V \alpha}{\tau} \exp{\left(\frac{-t}{\tau}\right)} \ \times  \int_0^t n_e(t')n_{ions}(t') \exp{\left(\frac{t'}{\tau}\right)}\ dt'.
\end{equation}
We define the normalized distribution $I_r$ as
\begin{equation}
I_r(t, \tau, T_R, \eta) = A \exp{\left(\frac{-t}{\tau}\right)} \ \times  \int_0^t n_e(t')n_{ions}(t') \exp{\left(\frac{t'}{\tau}\right)}\ dt',
\end{equation}
with $A$ a normalization factor. 

\acknowledgments
This work is part of the research program with project number FOM VP139, which is financed by the Netherlands Organisation for Scientific Research (NWO).
We gratefully acknowledge the technical support from to the mechanical, electrical and computing departments at Nikhef.
This work was carried out on the Dutch national e-infrastructure with the support of SURF Foundation.

\end{document}